\documentclass{aa}  
\usepackage{ulem}
\usepackage{graphicx}
\usepackage{txfonts}
\begin{document} 

   \title{Time evolution of snow regions and planet traps in an evolving protoplanetary disk.}


   \author{K. Bailli\'e
          \inst{1}
          \and
          S. Charnoz\inst{1}
          \and
          E. Pantin\inst{1}
          }

   \institute{Laboratoire AIM-LADP, Universit\'e Paris Diderot/CEA/CNRS,
                91191 Gif sur Yvette, France.\\
              \email{kevin.baillie@cea.fr}
             }

   \date{Received September 15, 2014; ...}

 
  \abstract
{Planet traps and snow lines are structures that may promote planetary formation in protoplanetary disks. They are very sensitive to the disk density and temperature structure. It is therefore necessary to follow the time evolution of the disk thermal structure throughout its viscous spreading. Since the snowlines are thought
to generate density and temperature bumps, it is important to take into account the disk opacity variations when the various
dust elements are sublimated.}
   {We track the time evolution of planet traps and snowlines in a viscously evolving protoplanetary disk using an opacity table that accounts for the composition of the dust material.
}
   {We coupled a dynamical and thermodynamical disk model with a temperature-dependent opacity table (that accounts for the sublimation of the main dust components) to investigate the formation and evolution of snowlines and planet traps during the first million years of disk evolution.}
   {Starting from a minimum mass solar nebula (MMSN), we find that the disk mid-plane temperature profile shows several plateaux (0.1-1 AU wide) at the different sublimation temperatures of the species
that make up the  dust. For water ice, the corresponding plateau can be larger than 1 AU, which means that this is a snow "region" rather than a snow "line". As a consequence, the surface density of solids in the snow region may increase gradually,  not abruptly. Several planet traps and desert regions appear naturally as a
result of abrupt local changes in the temperature and density profiles over the disk lifetime. These structures are mostly located at the edges of the temperature plateaux (surrounding the dust sublimation lines) and at the heat-transition barrier where the disk stellar heating and viscous heating are of the same magnitude (around 10 AU after 1 Myr).}
   {Several traps are identified: although a few appear to be transient, most of them slowly migrate along with the heat-transition barrier or the dust sublimation lines. These planet traps may temporarily favor the growth of planetary cores.}



\defcitealias{baillie14}{BC14}

   \maketitle
%

\section{Introduction} \label{intro}

Protoplanetary disk observations provide snapshots of the planetary formation processes that help us understand the physical characteristics of such disks. However, understanding planetary formation requires modeling the evolution and composition of protoplanetary disks. Considering the different physical states of the various components of a protoplanetary disk is therefore necessary to constrain favorable scenarios for planetary formation, growth, and migration, and solve the apparent inconsistency between the formation and migration timescales.

Since the discovery of the first exoplanet \citep{mayor95}, planetary formation scenarios have frequently been revisited. Even before \citet{charbonneau00} showed the gaseous giant nature of the observed exoplanets, \citet{pollack96} described how gaseous planets could form by accretion of gas on previously accumulated solid cores of a few times the mass of the Earth and estimated a typical planetary formation timescale around $10^{6-7}$ years (compatible with the typical disk lifetime of a few million years as inferred from observations by \citet{beckwith96} and \citet{hartmann98}). \citet{papaloizou99} also noted that hot Jupiters were unlikely to have formed in situ, therefore requiring some sort of planetary migration. Type I inward migration due to Lindblad resonances with the planet are well known from \citet{goldreich79}, \citet{arty93}, and \citet{ward97}, who estimated a migration timescale of about $10^{5}$ years for a typical Earth-mass planet in a minimum mass solar nebula \citep{weiden77, hayashi81}. \citet{kory93} noted that planets should therefore be lost by spiraling into their host star before they could actually grow. Early on, \citet{ward91} detailed the analytical expression of the corotation torque, before \citet{tanaka02} added a 3D analytical expression of the vortensity gradient horseshoe drag torque (tested numerically by \citet{bate03} and \citet{dangelo03}), and \citet{baruteau08} later completed this with the entropy gradient horseshoe drag torque. In the meantime, \citet{alibert05} and \citet{ida08} investigated the inconsistency between the timescales of planet formation and migration and noted that the mass-distance distribution of the exoplanets is inconsistent with the rapid inward migration of planetary cores. They noted that the type I inward migration should be slowed down by at least an order of magnitude to allow planets to form and avoid falling into the star. Part of the solution, investigated by \citet{hellary12}, would consist in accelerating the planet formation by considering a proper temperature treatment and a 3D disk model. However, this is not sufficient, and most of the efforts have been concentrated on slowing down the inward migration. Various models have been tested to achieve that goal. One of them is the model reported by \citet{terquem03},
who studied the effects of a regular magnetic field, while \citet{nelson04} focused on turbulent magnetism. \citet{kuchner02}, on the other hand, studied the consequences of an inner cavity and determined that such a hole could halt inward migration. \citet{jc05} found that self-shadowing in the disk can slightly decrease the migration rate. \citet{masset06b} described that a sharp positive surface mass density gradient could have similar consequences for the creation of planetary traps at the zero-torque radii. Subsequent efforts from \citet{paarm06,paarm08}, \citet{paarp08}, \citet{baruteau08}, \citet{kley08}, \citet{kley09}, and \citet{ayliffe10} investigated the possibility of slowing down and reversing the inward migration by considering more complete models involving proper radiative transfer and thermal consideration: this appeared to be possible for sufficiently low mass planets ($M_{P} \, <\, 40\, M_{\oplus}$). \citet{menou04} studied the effects of opacity transitions on the migration rate and showed that if some specific conditions are met, the migration could be stopped in a steady-state $\alpha$ disk. \citet{paarp09a} estimated that almost any positive surface mass density gradient could act as a protoplanet trap. Zero-torque radii were later analyzed by \citet{lyra10}, \citet{bitsch11}, and \citet{hasegawa112}: they appear to be very important in preventing the fall of the planets into their star, but also in creating a favorable zone for the interactions between planetesimals where they are able to combine. \citet{horn12} estimated that giant planet cores could form at convergent zero-torque radius for sub-$M_{\oplus}$ planets in 2-3 Myr.

There appears to be a strong correlation between the migration rate and the surface mass density and mid-plane temperature gradients that also strongly depend on the disk composition (gas-to-dust ratio, chemical abundances). Although the influence of the dust composition is not commonly used in recent numerical simulations of evolving protoplanetary disks, \citet{helling00} and \citet{semenov03} studied how opacities are affected by temperature. It therefore appears to be necessary to consider how the dust main component phases change in order to estimate the temperature more accurately. From reliable evolved disk radial profiles, we skip the planetary core formation stage and consider how a formed core dynamically interacts with the disk. The resonant torques that a planet exerts on a disk can be calculated from \citet{goldreich79}, \citet{ward88}, and \citet{arty93}, and we considered the refinements from \citet{paar11} to more accurately calculate the various contributions of the corotation torques. In their simulations, \citet{bitsch11} found a possible equilibrium radius of a planet with 20 Earth masses around 12.5 AU. \citet{hasegawa112} also analyzed the influence of the heat transition barrier on the migration torques. Whereas most of the previous work has applied the torque formulae to simplified semi-analytical density and temperature profiles \citep{hasegawa111,paar11}, to simple density prescriptions with a self-consistent 2D-temperature structure \citep{bitsch11,bitsch13}, or to steady-state accretion disk models \citep{bitsch14}, we intend to apply similar reasoning to more realistic disks obtained from numerically simulated viscous evolution rather than analytical steady state disks.

\citet{baillie14} (referred to hereafter as \citetalias{baillie14}) have shown that some of the features of viscous $\alpha$-model protoplanetary disks that are observed around forming stars can be retrieved numerically using a viscous evolution hydrodynamical code. These simulations confirmed the importance of jointly considering the dynamics, thermodynamics, and geometry of the disk. However, as the temperature affects the gas-to-dust ratio of the main components of the disk, the opacity of the disk is affected as well. Therefore, it is important to take into account a consistent composition of the disk when calculating its temperature. We here improve the numerical model of BC14 to consider these changes in the phases of the disk components. The thermal model includes both viscous heating and irradiation heating. We follow the evolution of an initial minimum mass solar nebula. The geometry (including the delimitations of shadowed regions) is calculated self-consistently, with the thermal structure obtained by semi-analytical radiative transfer calculations. The obtained density and temperature radial profiles show discontinuities compared to previous results from \citetalias{baillie14}. Density bumps mainly result in temperature irregularities such as bumps, troughs, and plateaux. The snow or sublimation lines are also enlarged. Using our density and temperature profiles, we compute the torque that the disk would exert on a planet embedded in the disk. The total torques (resulting from the Lindblad and corotation resonances) provide the direction of migration of the planetary core within the disk, allowing us to identify convergence and divergence regions, which are also called planetary traps and deserts.

Section \ref{methods} details how the hydrodynamical code of \citet{baillie14} is upgraded to consider variations of the dust composition with temperature. The density radial profiles resulting from the viscous evolution and the calculated thermodynamical and geometrical profiles of the disk are shown in Sect. \ref{res}. These radial profiles, and especially the mid-plane temperature, are analyzed in Sect. \ref{disc}, where we also discuss the influence of the composition of the disk and reconsider the definitions of snowline and sublimation line. Finally, Sect. \ref{torque} calculates the resonant torques of potential planetary embryos in the disk and follows their migration, defining planetary traps and deserts.


\section{Methods} \label{methods}
\subsection{Disk model}

We consider the same model of a viscous $\alpha$ disk \citep{shakura73} as was used in \citetalias{baillie14} and use the same terminology as in that paper. The turbulent viscosity is set to $\alpha_{\mathrm{visc}}= 10^{-2}$ as is commonly accepted for T Tauri star protoplanetary disks without deadzones \citep{fromang06}. The time evolution of the surface mass density is given by Eq. \ref{lb74} from \citet{lyndenbellpringle74},
\begin{equation}
\frac{\partial \Sigma(r,t)}{\partial t} = \frac{3}{r} \, \frac{\partial}{\partial r}\left(\sqrt{r} \, \frac{\partial}{\partial r} \left( \nu(r,t) \, \Sigma(r,t) \, \sqrt{r}\right) \right)
\label{lb74}
.\end{equation}

Similarly to BC14, we applied Eq. \ref{lb74} to a one-dimensional grid of masses that are logarithmically distributed in radius between $R_{*}$ and 1000 AU: each of these masses represents an annulus in the disk. We imposed that the flux at the inner edge cannot be directed outward. The inner mass flux gives the mass accretion rate of the disk. The temperature in the mid-plane, $T_{m}(r)$, results from the combination of viscous heating, stellar irradiation heating, and radiative cooling in the mid-plane.

The grazing angle $\alpha_{gr}(r)$ defines the angle at which the star sees the photosphere at a given radial location $r$. Comparisons between an imposed geometry following that prescription and a free geometry calculated along with a consistent temperature are shown in \citetalias{baillie14}, along with a discussion of the necessity of these geometric refinements. The grazing angle, which controls the amount of energy that the star provides to the disk photosphere, is related to the photosphere height $H_{ph}(r)$ through Eq. \ref{alphagr}:

\begin{equation}
\alpha_{gr}(r) = \arctan\left(\frac{dH_{ph}}{dr}(r)\right) - \arctan\left(\frac{H_{ph}(r)-0.4 R_{*}}{r}\right)
\label{alphagr}
.\end{equation}

A positive grazing angle at a given radius results in an irradiated photosphere at that location, while regions that are not irradiated are shadowed by inner regions. Using Eq. 18 from \citet{calvet91}, we calculated the temperature in the mid-plane, accounting for viscous heating, stellar irradiation, and radiative cooling. The viscous contribution depends on both the surface mass density obtained after temporal evolution and on the mid-plane temperature itself (through the viscosity):

\begin{equation}
F_{v}(r) = \frac{1}{2} \Sigma(r) \nu(r) \left( R \frac{\mathrm{d}\Omega}{\mathrm{d}r} \right)^{2} = \frac{9}{4} \Sigma(r) \nu(r) \Omega^{2}(r)
.\end{equation}

Therefore, we solved Eq. 18 from \citet{calvet91} numerically as an implicit equation on the mid-plane temperature. Considering a hydrostatic equilibrium, the vertical density distribution follows a Gaussian, and we can use Eq. A9 from \citet{dullemond01} to calculate the ratio $\chi$ of the photosphere height to the pressure scale height. The disk vertical density profile is therefore assumed to be the same as an isothermal vertical structure at the mid-plane temperature. This approximation is reasonable below a few pressure scale heights, where most of the disk mass is located. The opacities are also functions of the temperature, as detailed in Sect. \ref{realopa}. We can then estimate the corresponding presumed photosphere height $H_{ph}$ at each radial location and therefore access $\frac{dH_{ph}}{dr}$. Applying Eq. \ref{alphagr}, we can verify whether the presumed grazing angle has the required precision or if we should iterate on it. The impossibility of solving that problem for any positive value of the grazing angle results in a disk column that is not directly irradiated by the star, and therefore we removed the irradiation heating term from the mid-plane temperature equation.

Thus, the geometrical structure (photosphere and pressure heights) was determined jointly with the temperature by iterating numerically on the grazing angle value: the algorithm is thoroughly described in \citetalias{baillie14}.

\subsection{Realistic opacities}
\label{realopa}
While \citetalias{baillie14} did take into account both sources of heating, they only partially considered the thermal variations of the physical composition of the disk. Indeed, \citetalias{baillie14} considered that the dust grains invariably have the same opacity for any temperature below 1500 Kelvin (considered here to be the sublimation temperature of the silicate dust); and another constant opacity (a hundred times lower) for temperatures above 1500 K. In the present paper, we refine the model by considering a more elaborate model of opacities based on optical constants measured in laboratory experiments, accounting for the variations of dust composition as a function of the local temperature. The computation of the Rosseland mean opacities follows the procedure described by \citet{helling00} and \citet{semenov03}. We assumed that the dust grains are composed of a mixture of different elements: olivine silicate, iron, pyroxene, troilite, refractory and volatile organics, and water ice, with initial abundances (before volatiles sublimation) given in Table \ref{tempchgt}. The relative abundances of the opacity model are updated depending on the temperature of the medium and the sublimation temperatures (Table \ref{tempchgt}). We assumed for the dust grains a modified MRN size distribution \citep{pollack85, helling00,semenov03}, that is, grain sizes varying from 0.005 to 5 $\mu$m with a $-3.5$ power-law exponent size distribution. The absorption coefficients of the composite dust grains were computed using the Maxwell-Garnett mixing rule followed by the Mie theory. The tabulated values of the gas opacity computation are taken from \citet{helling00}.

\begin{figure}[htbp!]
\center
\includegraphics[width=\hsize]{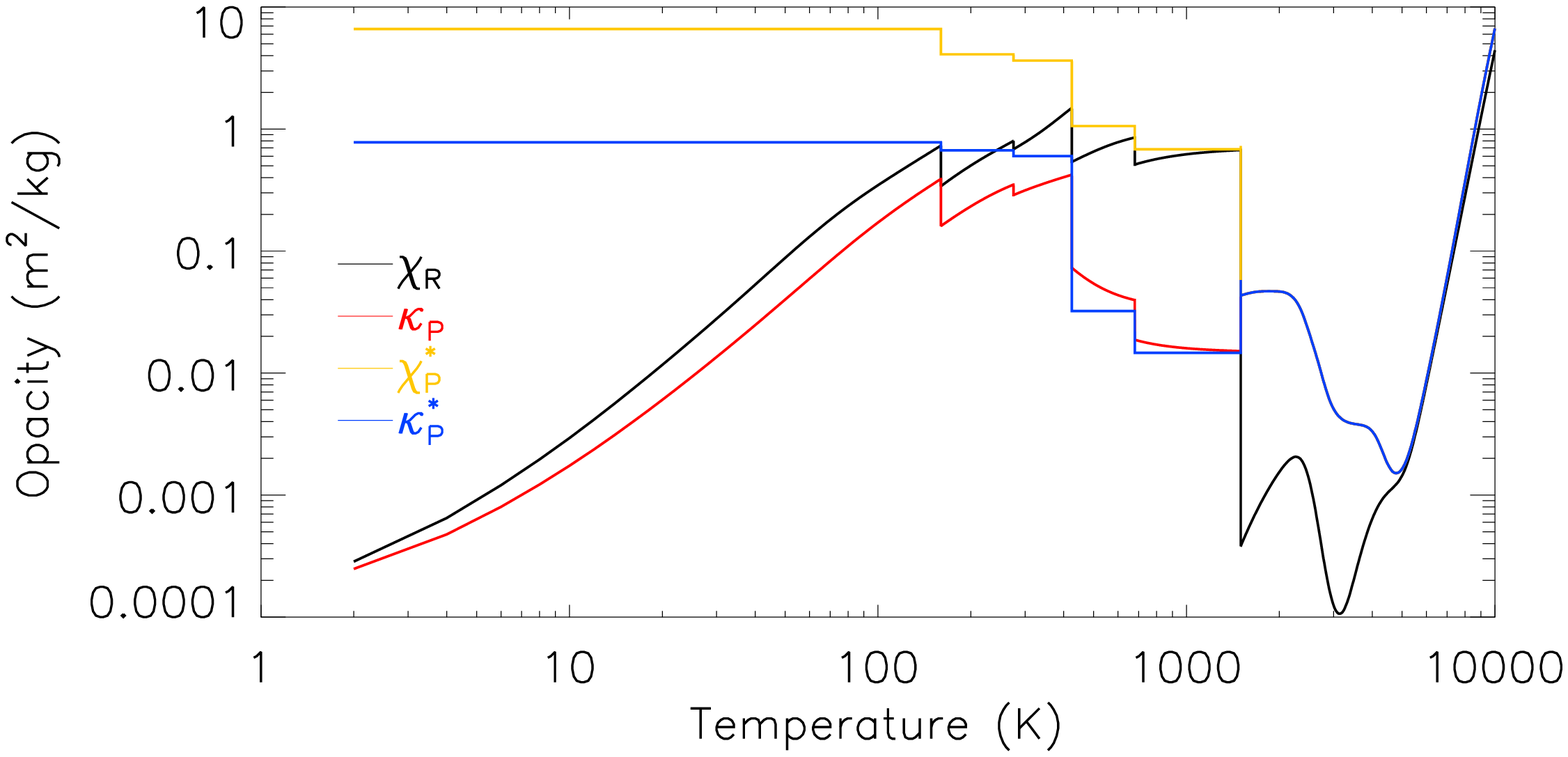}
\caption{Mean-opacity variations with local temperature. Black: Rosseland mean opacity in extinction. Red: Planck mean opacity in absorption. Yellow: Planck mean opacity in extinction at stellar irradiation temperature. Blue: Planck mean opacity in absorption at stellar irradiation temperature.}
\label{opa}
\end{figure}

Figure \ref{opa} presents the Rosseland ($\chi_{R}$) and Planck ($\kappa_{P}$) opacity variations with temperature. The Planck mean opacities at stellar effective temperature $T_{*} = 4000$ K in extinction ($\chi_{P}^{*}$) and absorption ($\kappa_{P}^{*}$) are also shown. These opacities are necessary in particular for calculating the irradiation heating as described in \citet{calvet91}, \citet{jc04}, and \citet{jc08}.

It appears that these opacities vary by several orders of magnitude over the concerned temperature range and that these variations are particularly steep near the sublimation temperature of the major disk gas components. These elements and their sublimation temperatures are presented in Table \ref{tempchgt}; sublimation temperatures are given by \citet{pollack94}, corresponding to gas densities of about $10^{-10} \, \mathrm{g.cm^{-3}}$.

\begin{table}
\begin{center}
$\begin{array}{c|c|c}
\mathrm{Elements} & \mathrm{Sublimation} & \mathrm{Relative}\\
& \mathrm{Temperature} & \mathrm{Abundances}\\
\hline
\mathrm{Water\, ice}            &       \mathrm{160\, K}        & 59.46 \, \% \\
\mathrm{Volatile\, Organics}    &       \mathrm{275\, K}        & 5.93 \, \% \\
\mathrm{Refractory\, Organics}  &       \mathrm{425\, K}        & 23.20 \, \% \\
\mathrm{Troilite\, (FeS)}               &       \mathrm{680\, K}      & 1.57 \, \% \\
\mathrm{Olivine}                &       \mathrm{1500\, K}       & 7.46 \, \% \\
\mathrm{Pyroxene}               &       \mathrm{1500\, K}       & 2.23 \, \% \\
\mathrm{Iron}                   &       \mathrm{1500\, K}       & 0.16 \, \% \\
\end{array}$
\end{center}
\caption{Sublimation temperatures and relative abundances that affect the disk gas opacity.}
\label{tempchgt}
\end{table}

\subsection{Testing the radial resolution}
\begin{figure}
\begin{center} $
\begin{array}{c}
\includegraphics[width=8cm, clip=true]{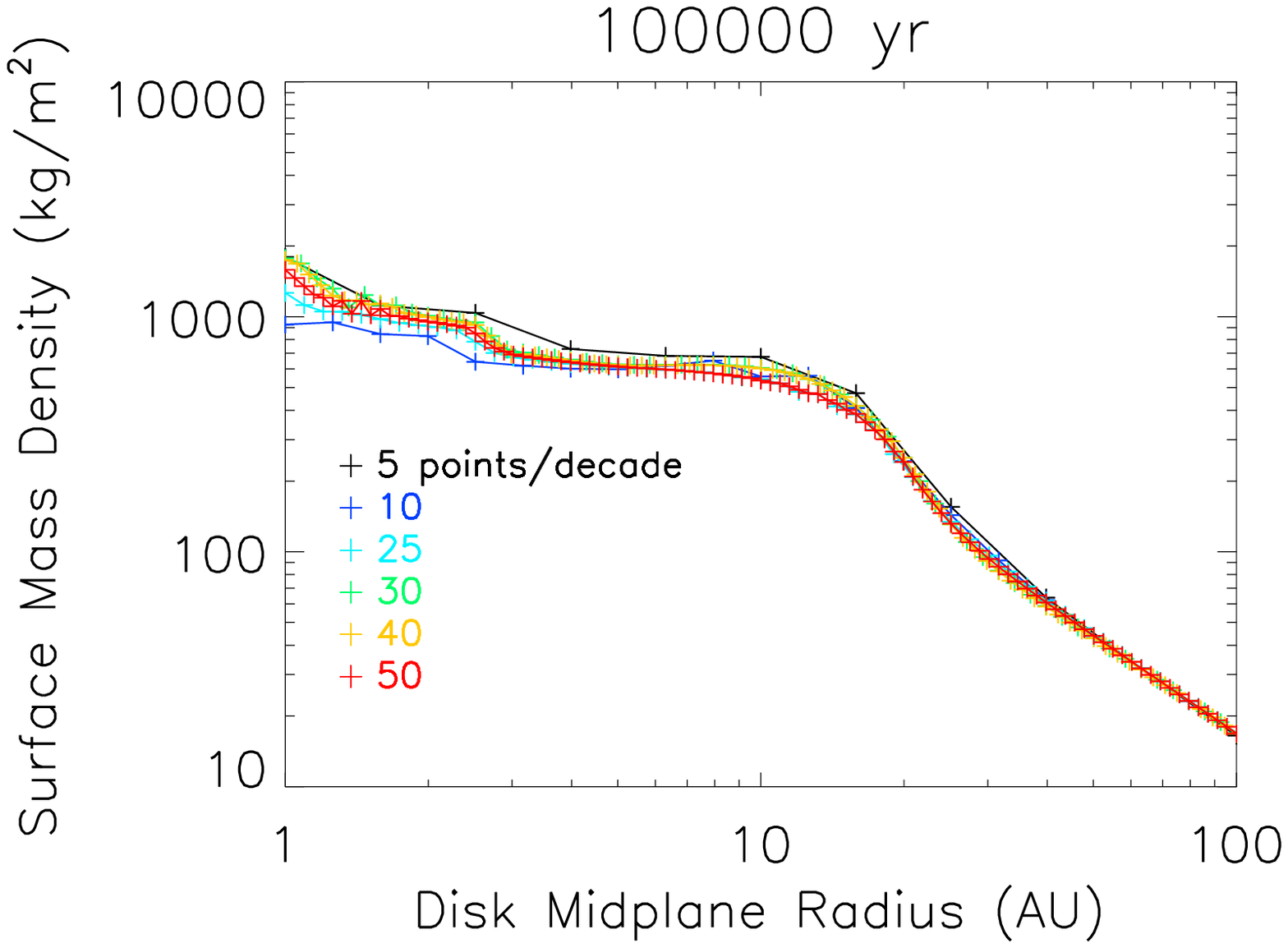}\\
\includegraphics[width=8cm, clip=true]{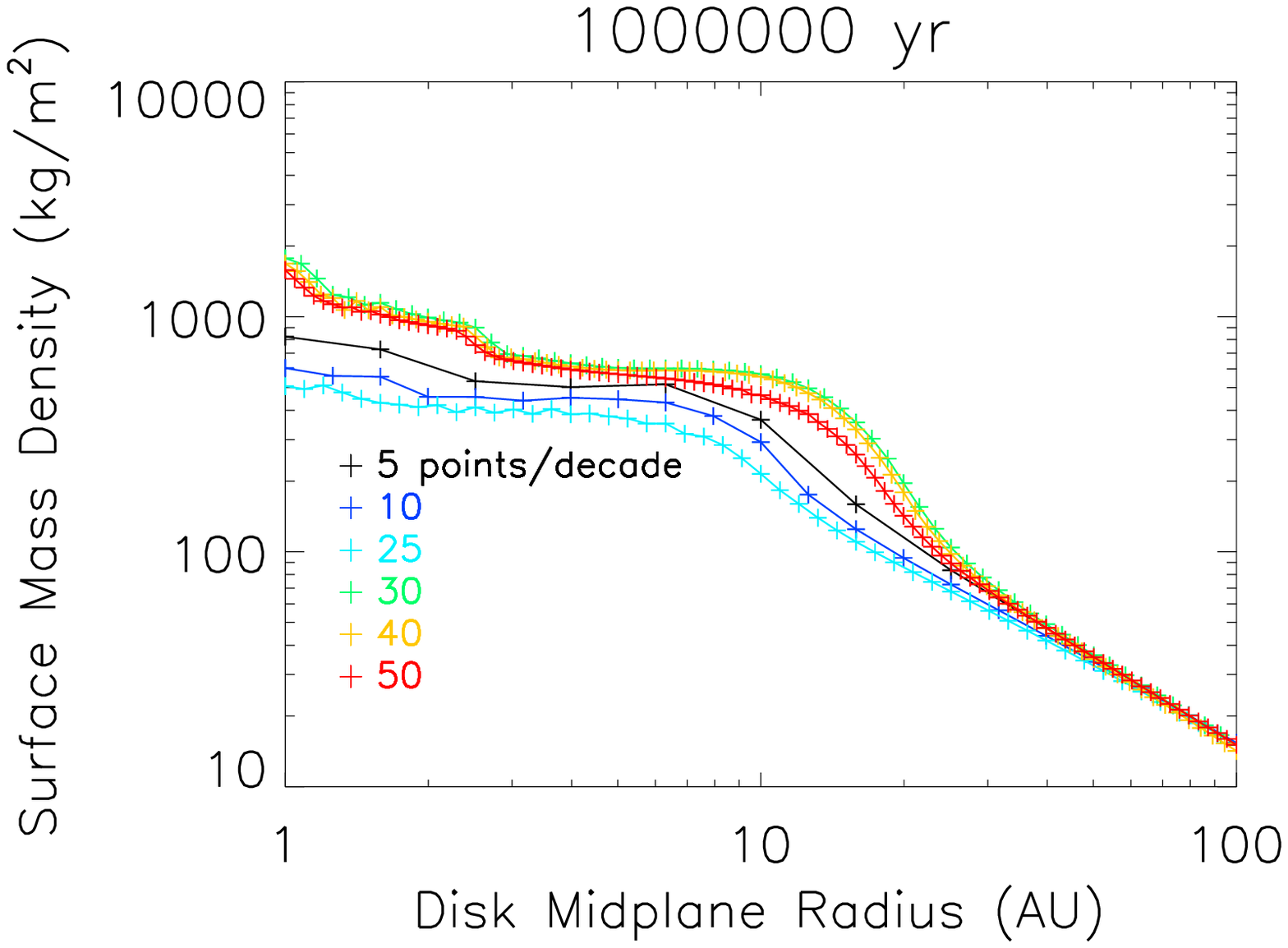}\\
\end{array} $
\end{center}
\caption{Compared evolution of the surface-mass density after 100,000 years (upper panel) and 1 million years of evolution (lower panel) of an initial MMSN for various resolutions from 5 to 50 points per decade.}
\label{rescomp}
\end{figure}

Our evolution code requires a double check both on the temporal and on the spatial resolution that the structure of our code (density time evolution precedes geometry and temperature calculation and so on for each iteration) allows us to treat sequentially. First, the timestep is adjusted to limit the mass transfers to 1\% of the available mass in each bin of the simulation, and then the radial resolution can be controlled by ensuring that the surface mass density profiles are consistent over 1 million years for different radial resolutions. Figure \ref{rescomp} shows the radial density profiles between 1 and 100 AU after 100,000 years and 1 million years for a range of radial resolutions extending from 5 points per decade to 50 points per decade. Stronger differences around 1 AU are due to the boundary conditions of the test numerical simulations for which the inner edge of the disk was considered to be at 1 AU for the purpose of the comparison: as the computation time increases exponentially, simulations using the highest number of points per decade could only be run over a few decades. We note that the surface mass density profiles are very similar until 100,000 years, after which the differences are more obvious. However, the density profiles remain quite close and the only consequence of the difference seems to be a delay in the viscous evolution for the simulations with 30 and more points per decade.

Therefore, we estimate that a disk evolution with a good radial resolution (e.g., 40 points per decade as in the rest of the paper) can be approximated from interpolating the density radial profile of a disk generated by a less radially resolved simulation (5 points per decade) for an accessible computation time. The thermal and geometric structure can then be calculated based on the resampled density structure of the disk.

Assuming a temperature profile following a power law in $T \propto r^{-0.5}$ and a temperature resolution of 0.01 K, we can analytically estimate the highest possible radial resolution for two consecutive bins to have temperatures that differ by at least that temperature precision: this optimal resolution is around 100 points per decade. To keep a safety margin (in particular to allow for the temperature fluctuations due to the opacities), we chose to resample our disks with 40 points per decade, which provides a good radial precision and allows a good thermal and geometrical precision.


\section{Results} \label{res}
We followed the evolution of an initial minimum mass solar nebula \citep{weiden77} with the scaling from \citet{hayashi81}:

\begin{equation}
\label{eqmmsn}
\Sigma (r) = 17,000 \left(\frac{r}{1 \, \mathrm{AU}}\right)^{-3/2} \mathrm{kg\cdot m^{-2}}
.\end{equation}

The central star was a classical T Tauri type young star with constant $M_{*} = 1 \, M_{\odot}$, $R_{*} = 3 \, R_{\odot}$, $T_{*} = 4000 \, \mathrm{K}$ and $\mathcal{L}_{*} = 4 \pi \, R_{*}^{2} \, \sigma_{B} \, T_{*}^{4}$ throughout the simulation.

The choice of the minimum mass solar nebula (MMSN) is motivated by several arguments. First, \citetalias{baillie14} investigated a diversity of initial disk conditions (total disk mass and radial distribution of the initial angular momentum) and found that these numerical simulations converged to similar steady states (characterized by a uniform mass flux and an asymptotic surface mass density distribution in $r^{-1}$), although in slightly different timescales. In addition, \citet{vorobyov07} showed that the MMSN density profile (following a power-law in $r^{-1.5}$) was consistent with an intermediate stage of an evolving protoplanetary disk under self-regulated gravitational accretion. Therefore, the MMSN profile makes an initial profile as reasonable as any other snapshot that could have been taken in the disk evolution. Finally, this fiducial case makes sense because we can better compare our results with previous studies.

\subsection{Time evolution}
\label{evol}
\begin{figure}[htbp!]
\center
\includegraphics[width=\hsize]{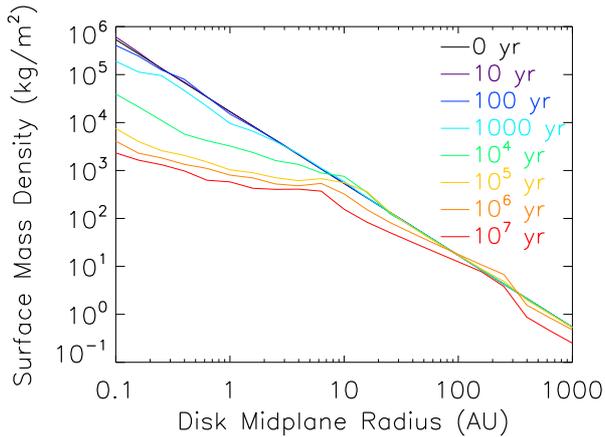}
\caption{Surface mass density radial profile evolution for an initial minimum mass solar nebula in the case of a self-consistently calculated geometry with a full continuous model of opacities. Simulations are resampled to 40 points per decade.}
\label{sigma}
\end{figure}

\begin{figure}[htbp!]
\center
\includegraphics[width=\hsize]{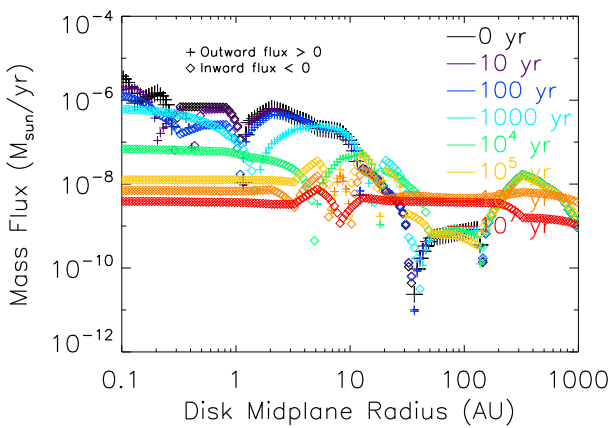}
\caption{Mass flux radial profile evolution for an initial minimum mass solar nebula in the case of a self-consistently calculated geometry with a full continuous model of opacities.}
\label{fluxb}
\end{figure}

Figure \ref{sigma} presents the evolution of such a disk over 10 million years. Although the protoplanetary disk gas is believed to photo-evaporate in a few million years \citep{font04, alexander07, alexander09, owen10}, we let our numerical simulations extend over longer times to reach a stationary steady-state, which is
characterized by a uniform mass flux (within one order of magnitude maximum), as seen in Fig. \ref{fluxb}.

As \citetalias{baillie14} detailed, the disk starts to spread viscously outward before beginning to accrete onto the central star after a few thousand years. Figure \ref{sigma} shows that the surface mass density radial profile can be modeled as a power
law by segments. While two segments seem to be sufficient until 10,000 years, longer evolution times require at least one more segment. In the early times, the single connexion point between the two power-laws connects an inner region that has already evolved toward a steady-state power-law radial profile for this region, and an outer region that is much closer to the initial state. We therefore call that radial location the relaxation radius. After 10,000 years, the (inner) relaxation radius is located between 7 and 10 AU. After 1 million years, a secondary "knee" appears around 250 AU. In the intermediate region (10-250 AU), the viscous evolution appears weaker than in the simulations
of \citetalias{baillie14} with a simpler opacity model. At 1 Myr, the surface mass density profile can be modeled as a power
law between 10 and 250 AU with $\Sigma(r) \propto r^{-1.2}$, and at 10 Myr, we can approximate the density profile by $\Sigma(r) \propto r^{-1.1}$. In this region, and evolved disks, the power-law indices are comparable with those from \citetalias{baillie14} and with those observed by \citet{isella09} in the Taurus region and \citet{andrews09} and \citet{andrews10} in the Ophiuchus region.

Figure \ref{fluxb} shows the time evolution of the mass flux as a function of the radial distance. We note that these radial profiles present bumps that were not visible in \citetalias{baillie14}. Although the directions and amplitudes remain globally identical, some irregularities appear; for example, in the 7-10 AU region where the 1 Myr flux profile is briefly inverted and the 10 Myr profile is no longer uniform over that range (the flux varies by almost a  factor 10).

\subsection{Thermal evolution}
\label{Tevol}

\begin{figure}[htbp!]
\center
\includegraphics[width=\hsize]{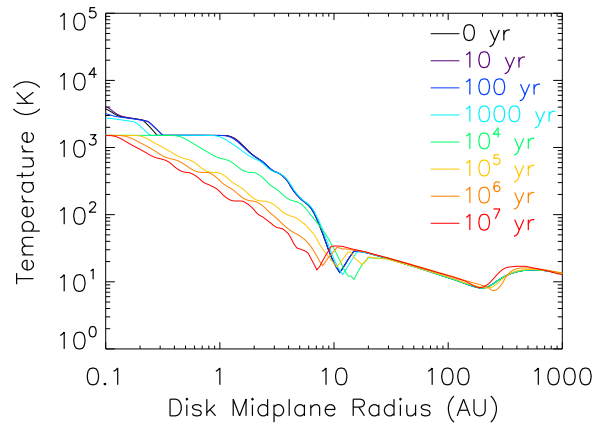}
\caption{Evolution of the mid-plane temperature as a function of the radial distance for an initial minimum mass solar nebula in the case of a self-consistently calculated geometry with a full continuous model of opacities.}
\label{Tm}
\end{figure}

The thermal evolution of the disk is presented in Fig. \ref{Tm}. The mid-plane temperature is calculated in a similar way to \citetalias{baillie14}. However, the calculation of the mid-plane temperature takes into account a composition of the dust that is consistent with the temperature (i.e., where sublimated species have been removed from the opacity calculation according to Table \ref{tempchgt}) while iterating over the possible grazing angles and mid-plane temperatures: the temperature obtained after the algorithm converged therefore validates the consistent composition. As described in Sect. \ref{methods}, the geometric and thermal structure are recalculated over 40 points per decade.

As in \citetalias{baillie14}, the mid-plane temperature presents irregular features that did not appear in the radial profiles obtained with the simpler opacity model of \citetalias{baillie14}. In particular, we note the temperature plateaux at the change of phase temperatures of the dust components (see Table \ref{tempchgt}). As the disk evolves, the sublimation plateaux drift inward: the silicate sublimation plateau is inside 0.2 AU after 1 Myr while it was originally between 0.3 and 1.3 AU. In addition to these plateaux, we note the troughs around 10 AU and 250 AU, which drift inward as the disk evolves. These features are analyzed more thoroughly in Sect. \ref{tempprofile}. However, it appears that the temperature in the region between these features can be modeled as a power law, with $T(r) \propto r^{-0.47}$ beyond 1 Myr, which recalls the usual approximation of \citet{chiang97} or \citet{dullemond01}, according to which $T(r) \propto r^{-1/2}$.

\subsection{Geometry evolution}
\label{geoevol}

\begin{figure}[htbp!]
\center
\includegraphics[width=\hsize]{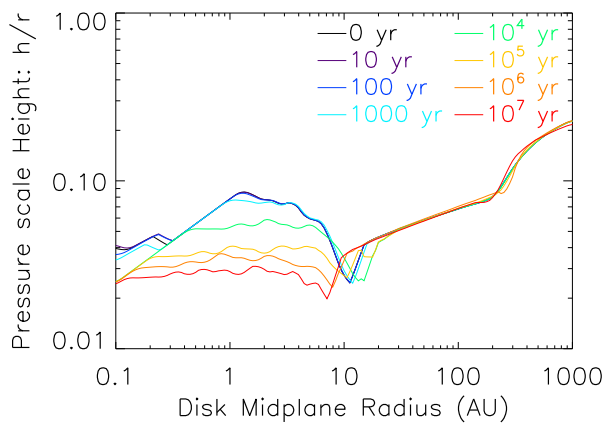}
\caption{Evolution of the pressure scale height as a function of the radial distance for an initial minimum mass solar nebula in the case of a self-consistently calculated geometry with a full continuous model of opacities.}
\label{hcgm}
\end{figure}

As the geometry of the disk is calculated jointly and self-consistently with the thermodynamical structure, irregular features can also be expected in the height radial profiles, as shown in Fig. \ref{hcgm}: the pressure scale height shows radial gradient discontinuities similar to those observed in Fig. \ref{Tm}. Similarly, it is possible to approximate the pressure scale height by an almost constant power law between 20 AU and 250 AU: $h_{\mathrm{pressure}} =  r^{1.26}$ after 1 Myr and $h_{\mathrm{pressure}} =  r^{1.28}$ at 10 Myr. These index values are very close to the $9/7$ index suggested by \citet{chiang97} in the case of a passive disk for which the photosphere height (height at which the line-of-sight optical depth equals 1) would be directly proportional to the pressure scale height. It appears then that the approximation
of $h_{\mathrm{pressure}} \propto r^{9/7}$ can only be valid in the region between 10 and 250 AU, where the disk photosphere is irradiated and the opacity varies smoothly with the temperature (i.e., where the temperature is lower than the sublimation temperature of the main components, as listed in Table \ref{tempchgt}).

\begin{figure}[htbp!]
\center
\includegraphics[width=\hsize]{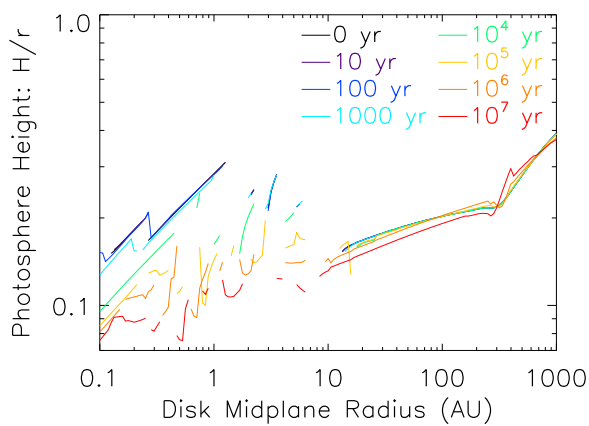}
\caption{Evolution of the photosphere height as a function of the radial distance for an initial minimum mass solar nebula in the case of a self-consistently calculated geometry with a full continuous model of opacities.}
\label{Hm}
\end{figure}

The radial profile of the photosphere height (Fig. \ref{Hm}) reveals a zone inside of 10 AU where the disk photosphere is irregularly irradiated, while it is entirely irradiated in the outer regions. Between 20 and 250 AU, the pressure scale height can be approximated by a power law: $H_{\mathrm{photo}} \propto  r^{1.11}$ after 100,000 yr and $h_{\mathrm{pressure}} \propto  r^{1.14}$ at 1 Myr and 10 Myr. These index values are reminiscent of those expected by \citet{kenyon87} around $9/8$ and obtained by the numerical simulations of \citetalias{baillie14}.

\begin{figure}[htbp!]
\center
\includegraphics[width=\hsize]{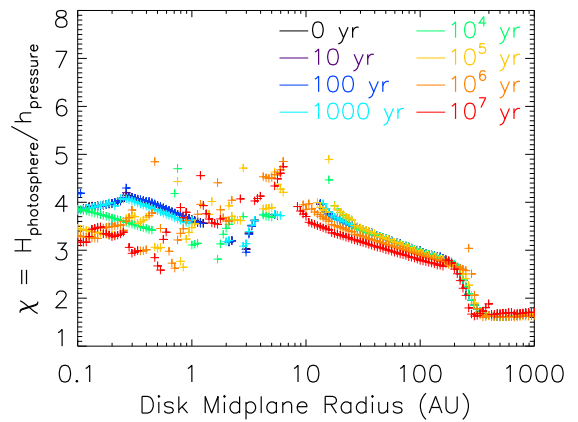}
\caption{Evolution of the pressure scale height to photosphere height ratio as a function of the radial distance for an initial minimum mass solar nebula in the case of a self-consistently calculated geometry with a full continuous model of opacities. Missing points are due to the regions of the disk that are not directly in the stellar line of sight at a given location and evolution time.}
\label{khim}
\end{figure}

\begin{figure}[htbp!]
\center
\includegraphics[width=\hsize]{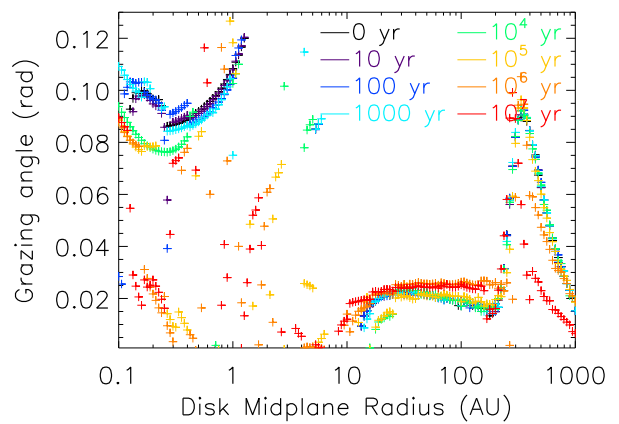}
\caption{Evolution of the grazing angle as a function of the radial distance for an initial minimum mass solar nebula in the case of a self-consistently calculated geometry with a full continuous model of opacities. Missing points are due to the regions of the disk that are not directly in the stellar line of sight at a given location and evolution time.}
\label{alphagrm}
\end{figure}

The evolution of the photosphere height over pressure scale height ratio $\chi$ is displayed in Fig. \ref{khim}. As predicted by \citet{dullemond01}, the values are in the range from 1.5 to 6. However, unlike the usual approximation ($\chi = 4$) suggested by \citet{chiang97}, this ratio is neither constant nor uniform. This $\chi$ profile, like the grazing angle $\alpha$ profile (Fig. \ref{alphagrm}) show similar gradient discontinuities as in the surface mass density or temperature radial profiles. The non-irradiated zones also appear quite obvious in these figures because they correspond to locations where $\chi$ and the grazing angle are not defined.

\section{Discussion} \label{disc}

\subsection{Influence of the disk composition on density and temperature local gradients}

By comparing these results with the simulations of \citetalias{baillie14}, we note how important it is to take the physical composition of the dust into account by using a complete model of opacities that varies with temperature. The main difference in the surface mass density radial profiles resides in the secondary knee that appears at 1 Myr around 250 AU. Earlier conclusions from \citetalias{baillie14} remain valid: the surface-mass density profile evolves toward a shallower profile, first in the inner regions and then increasingly farther away from the star. However, when the relaxation radius reaches 250 AU, the profile is then fragmented into three power-law segments. This creates a discontinuity in the density gradient that may strongly influence the exchange of angular momentum with possible planetary cores within the disk (see Sect. \ref{torque}). In addition to that, we note that the mass flux may reverse in the middle of the disk until as late as 1 Myr. These density bumps obviously have consequences for the thermodynamical and geometrical structures. The mid-plane temperature and grazing angle radial profiles do in fact present new features, particularly at the radii of these discontinuities, when the physical composition of the disk is properly treated. Temperature plateaux are other important consequences as they may affect, for instance, the snowline position (see Sect. \ref{snowline}).

\subsection{Temperature radial profile analysis}
\label{tempprofile}

\begin{figure}[htbp!]
\center
\includegraphics[width=\hsize]{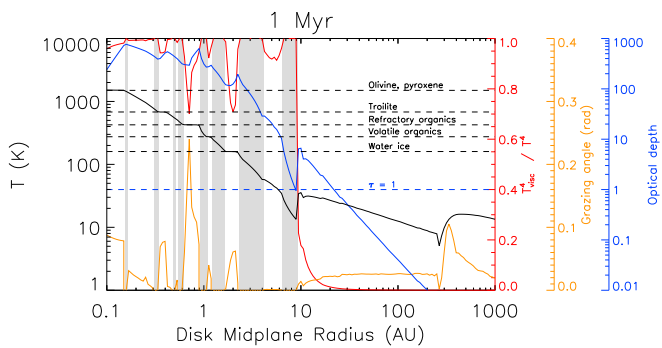}
\caption{Mid-plane temperature radial profile (black) after 1 million years of evolution of a minimum mass solar nebula with a self-consistently calculated geometry and a full continuous model of opacities. Disk shadowed regions are displayed in gray. The ratio of the viscous heating contribution over the total heating (viscous heating rate) is presented in red, the grazing angle radial profile in yellow, and the optical depth radial profile in blue.}
\label{profilT}
\end{figure}

Figure \ref{profilT} details the temperature structure in the mid-plane of the protoplanetary disk after 1 Myr of evolution. As well as the mid-plane temperature, Fig. \ref{profilT} also displays the grazing angle and optical depth radial profiles, and the distribution of the heating sources.

We first note is that the black curve shows temperature plateaux that coincide with the sublimation temperatures listed in Table \ref{tempchgt}. These plateaux are not flat, but present variations of a few Kelvin over radial regions that may be up to 1 AU wide. As Fig. \ref{opa} showed, opacities may vary quite strongly with temperature around these changes of phases. Thus, the temperature drop induced by a local surface mass density variation can be compensated for by an opacity variation over a few Kelvin to provide the equivalent heat and maintain a quasi-constant temperature. In these regions, the irradiation heating effectively has a more important contribution than outside, where it tends to become negligible compared to the viscous heating. This is confirmed by the red curve, which represents the ratio of the viscous contribution over the total heating received by the disk (viscous heating rate): the troughs at 0.1, 0.4, 0.7, 1.2, and 2 AU coincide with the temperature plateaux. This transition is called heat transition barrier by \citet{hasegawa112}. In addition, we note that the silicate sublimation plateau migrates inward as the disk evolves, and after 1 Myr, it is located at the disk inner edge. We interpret the inner disk physical structure as follows:
\begin{itemize}
\item The surface mass density is generally a decreasing function of the radial distance.
\item When a phase-transition temperature is reached, the opacity suddenly increases because of new condensed species, even though the temperature remains stable over a few fractions of an AU.
\item The viscosity $\nu = \alpha_{\mathrm{visc}} \, c_{s} \, h_{\mathrm{pressure}}$ therefore follows the temperature stability. As the density and the angular velocity decrease outward, the quantity of viscous heating decreases as well.
\item To maintain the temperature, the irradiation heating tries to compensate for the viscous heating loss, looking for an optimal grazing angle that can maximize the irradiation heating from the star.
\item At some radii, the disk photosphere geometry is such that the irradiation heating can no longer compensate for the loss in viscous heating. The grazing angle then drops until the disk photosphere is not irradiated anymore. The viscous heating remains the only source of heating, resulting in a shadowed region.
\item The temperature decreases again because of the lack of irradiation heating. However, the opacity variation is smoother now that we are no longer at the phase transition temperature. The viscosity now decreases, reinforcing the decrease in viscous heating. In the meantime, the opacity is now much higher than it was before the change of phase.
\end{itemize}

In addition to these plateaux, the mid-plane temperature radial profile shows two important troughs with an amplitude drop of about 10 Kelvin. The first one, located around 10 AU, coincides with the limit where the dominating source of heating changes: viscous heating is clearly stronger below 10 AU, while stellar irradiation heating dominates outside that limit (red curve from Fig. \ref{profilT}). The second trough, around 250 AU, generates a drop in temperature from 20 to 8 Kelvin. At these temperatures, the opacity varies by two orders of magnitude over a temperature range of 20 Kelvin (Fig. \ref{opa}): the grazing angle drops and the geometry tends to shadowing. The temperature bump disappears in simpler opacity models \citepalias{baillie14}. In addition, it remains present (although less pronounced) in earlier evolution ages.

Furthermore, we note that in the irradiation-dominated region, the disk is permanently irradiated, which is not the case in the inner regions (see gray bands in Fig. \ref{profilT}). These shadowed regions coincide with a drop in the grazing angle (see yellow curve in Fig. \ref{profilT}). Moreover, all the temperature plateaux (all are located below 10 AU) correspond to irradiated zones, while the outer edges of the plateaux (where the temperature starts dropping again) trigger shadowed regions, as detailed above. The region between 20 and 250 AU matches the two-layer model of \citet{chiang97} well because the temperature is clearly below the dust sublimation temperatures, and therefore in a temperature domain where the opacity varies so smoothly that it can approximated by a constant, allowing the retrieval of the simulation results from \citetalias{baillie14}. 

Finally, the stellar line-of-sight optical depth radial profile (blue curve in Fig. \ref{profilT}) shows that the disk becomes optically thin beyond 20 AU. However, the optical thickness drops again at the heat transition barrier (around 10 AU at 1 Myr).

\subsection{Snow region}
\label{snowline}
\begin{figure}[htbp!]
\center
\includegraphics[width=\hsize]{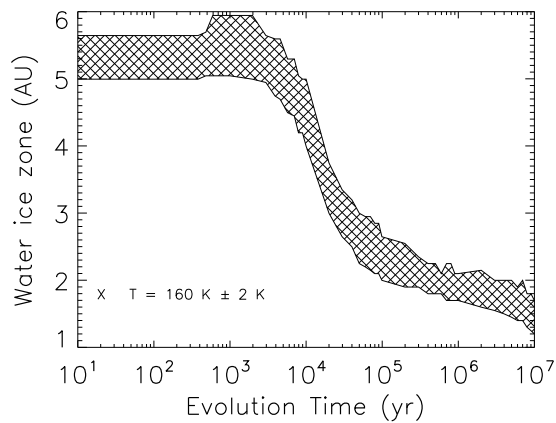}
\caption{Time evolution of the snow region (mid-plane radial location for which the temperature coincides with the water-ice condensation temperature $\pm$ 2 K).}
\label{snowzone}
\end{figure}

While the snowline is defined as the radius at which the temperature is equal to the water ice condensation temperature (160 K), the fact that we now observe a plateau around that temperature shows that we should rather talk about a snow region than a snowline. Figure \ref{snowzone} presents the evolution of the inner and outer edges of the plateau around the water ice condensation temperature. Given the precision on the temperature, and in the opacity model, we chose to define the plateau temperature range as $160 \, \mathrm{K} \, \pm \, 2 \, \mathrm{K}$. The snow region can be as wide as 1 AU at early ages and migrates inward until it stabilizes below 2 AU in a few Myr.

\begin{figure}[htbp!]
\center
\includegraphics[width=\hsize]{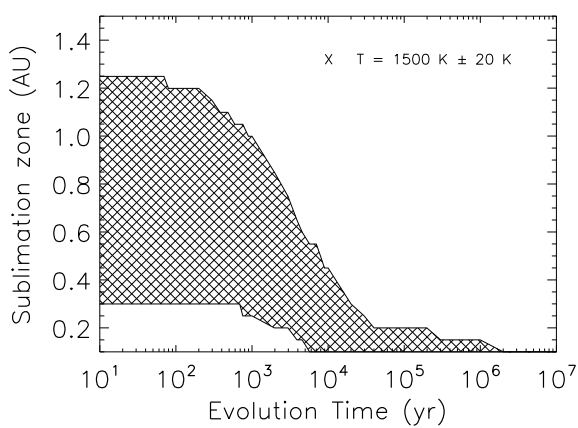}
\caption{Time evolution of the silicates sublimation zone (mid-plane radial location for which the temperature coincides with the silicate sublimation temperature $\pm$ 20 K).}
\label{sublzone}
\end{figure}

Similarly, we define the silicate sublimation region in place of the silicate sublimation line at $1500 \, \mathrm{K} \, \pm \, 20 \, \mathrm{K,}$ and we present the evolution of its edges in Fig. \ref{sublzone}. The sublimation region initially spreads from 0.3 to 1.3 AU. It also migrates inward and seems to reach the inner edge of the disk in a few Myr.

It appears that the zone of the mid-plane in which the sublimation of an element can occur is AU-wide instead of the expected sharp frontier usually called snow or sublimation line. At this temperature, the corresponding element is sublimated, which affects the medium opacity and therefore the heating received by the disk at this location. This heating variation is compensated for by a grazing angle variation that maintains a smoother continuity of the temperature and gas-to-dust ratio. The origin of the plateau is therefore related to the partial and gradual sublimation of the element that is sublimated at this temperature: the gas-to-dust ratio of that element is close to 0 at the outer edge of the plateau while it increases inward up to 1 at the inner edge. The other elements gas-to-dust ratios are mostly unaffected at this temperature. Partial sublimation could occur layer by layer on dust particles or on a vertical scale (the column height of the sublimated material increasing inward).

The changes of phase can now occur in much wider regions than the previous snowlines. This may favor smooth variations in physical compositions across the disk as the planetary embryo migrates inward across the snow or sublimation region. This may influence the chemical models, for example, those that try to explain the abundance of carbon in the so-called carbon-rich planets: \citet{fortney10} and \citet{madhu11} suggested that, at equilibrium, all the available O should go into organics, whereas \citet{venot12} described a non-equilibrium chemistry to model the C/O ratio. Therefore, the shallow variations of the temperature profile may favor equilibrium chemistry.



\section{Consequences on planet traps and deserts}
\label{torque}
The results reported above clearly show that rapid variations of temperature and density occur in the disk in the transition region. It is thought that such a transition may potentially create planets traps and deserts. This has been explored previously in static-disk models, but never in dynamically evolving disks. Thus in the following we compute the torque that the disk would exert on a putative planet. In particular we explore how planet traps appear, move, and disappear in the disk.

Assuming that it has already formed, a planetary embryo exchanges angular momentum with the disk \citep{goldreich79,ward88,arty93,jc05}. These exchanges are due to the resonances excited by the planetary embryo in the disk (Lindblad resonances caused by the action of the induced spiral arms, and corotation resonances). Thus, the planet exerts a torque on the disk and therefore the disk exerts an opposite torque on the planet. We calculate these torques in the case of the evolved disk described in the previous section. We then study their effects on potential planetary embryos. We assume here that the disk structure is not modified by the planet.



\subsection{Lindblad torques}
A given perturber, such as the planet in our disk, excites Lindblad resonances of multiple orders \citep{goldreich79,goldreich80}. Using a two-dimensional approximation, considering laminar disks, a planet on a circular orbit, ignoring the disk self-gravity and assuming thermal equilibrium, \citet{paarp08} were able to derive the following formula for the total Lindblad torque exerted by the disk over the planet:

\begin{equation}
\label{gammalin}
\Gamma_{\mathrm{Lindblad}} = - \frac{\Gamma_{0}(r_{P})}{\gamma} \left(2.5 \, - 1.7 \frac{\partial \ln T}{\partial \ln r}(r_{P}) + 0.1 \frac{\partial \ln \Sigma}{\partial \ln r}(r_{P}) \right)
,\end{equation}

with $\gamma = 1.4$, the adiabatic index,\\
$\Gamma_{0}(r_{P}) = \left(\frac{q}{h}\right)^{2} \, \Sigma(r_{P}) \, r_{P}^{4} \, \left(\Omega(r_{P})\right)^{2}$,\\
$h=\frac{h_{\mathrm{press}}(r_{P})}{r_{P}}$,\\
and $\Omega(r_{P})$ the Keplerian angular velocity at the planet position in the disk.

Although we just showed in the previous sections that the disk can present radial variation in density and temperature gradients, the torque expressions are still evaluated at the planet radial location. \citet{ward97}, \citet{hasegawa111,hasegawa112}, and \citet{masset11} (Appendix B) developed fully analytic torque expressions that could account for the local gradients at each resonance location rather than just considering the gradients at the planet radius. However, these formulas could not retrieve the amplitude of the expression derived in \citet{paarp08}, which was benchmarked against numerical simulations. In addition, more subtle expressions involving the second derivative of the temperature derived by \citet{masset11} (Eq. 79) were only tested in an isothermal disk. To be consistent with the corotation torque expressions from \citet{paar11}, we therefore estimated the total Lindblad torque using the expression of \citet{paarp08} (our Eq. \ref{gammalin}).

We note that the Lindblad torque presents a stronger dependence on the temperature gradient than on the density gradient, which reflects the effect of the pressure buffer described in \citet{ward97}. Following a similar process as in \citet{bitsch11} and \citet{bitsch13,bitsch14}, we can use here the results of Sect. \ref{res}, which provide the density and temperature of an evolved disk at a given date, therefore gaining in consistency as the temperature is not set to a power law and is calculated jointly with the geometry of the disk from the density resulting of the viscous evolution.


\subsection{Corotation torques}
Corotation resonances are known to exert complicated torques that include linear and nonlinear parts. \citet{paarp09b} showed that the corotation torques are generally nonlinear in the usual range of viscosity ($\alpha_{\mathrm{visc}} < 0.1$). The nonlinear contribution, due to the horseshoe drag \citep{ward91} caused by the interaction between the planet and the fluid element moving in its vicinity, is also known for having two possible origins: barotropic, intially formalized by \citet{tanaka02}, and entropic, detailed by \citet{baruteau08}.

Concerning the horseshoe drag, \citet{paar11} described the density perturbation generated by the corotation resonances and provided expressions for both the entropy and vortensity (or barotropic) contributions. Assuming a gravitational softening $b=0.4 h_{\mathrm{press}}$, \citet{bitsch11} and \citet{bitsch14} summarized these expressions to obtain the following contributing torques:

\begin{eqnarray}
\label{gammahsentro} \Gamma_{\mathrm{hs,entro}} &=& - \frac{\Gamma_{0}(r_{P})}{\gamma^{2}} \, 7.9 \, \left(-\frac{\partial \ln T}{\partial \ln r}(r_{P}) + (\gamma-1) \frac{\partial \ln \Sigma}{\partial \ln r}(r_{P}) \right)\\
\label{gammahsbaro} \Gamma_{\mathrm{hs,baro}} &=& - \frac{\Gamma_{0}(r_{P})}{\gamma} 1.1 \left(\frac{\partial \ln \Sigma}{\partial \ln r}(r_{P}) + \frac{3}{2}\right)
.\end{eqnarray}

\citet{paar11} showed that in the absence of saturation of the linear contributions, the total corotation torque can be defined as

\begin{equation}
\label{gammacor}
\Gamma_{\mathrm{corotation}} = \Gamma_{\mathrm{hs,baro}} + \Gamma_{\mathrm{hs,entro}}
.\end{equation}

It appears that this unsaturated corotation torque strongly depends on the temperature and surface mass density gradients. It also
scales with $M_{\mathrm{P}}^2$, as does the Lindblad torque. However, \citet{paarp09a} showed that given the viscous, diffusive, and libration timescales, the linear effects of the corotation torques can be saturated for some viscosities and some planet masses. For our disk that evolved for 1 Myr, the viscosity range compared to Fig. 14 from \citet{paarp09a} suggests that saturation cannot be neglected for planetary masses higher than $6 M_{\mathrm{\oplus}}$. \citet{paar11} defined weight functions for the partial saturation of the corotation torque. These functions vary with the half-width of the horseshoe, which depends on the mass of the planet. Appendix A of \citet{bitsch11} summarized this method and added correcting factors. We used a similar torque calculation, which is necessary to take into account the variations with the planet mass.

\subsection{Planetary core migration}

\subsubsection{Migration direction}
In the present exercise, we set the planet mass to a typical giant planet core mass $M_{P} = 10 \, M_{\mathrm{\oplus}}$ and varied $r_{P}$, the initial planet location. The total torque exerted by the disk over the planet is given by

\begin{equation}
\label{gammatot}
\Gamma_{\mathrm{tot}} = \Gamma_{\mathrm{Lindblad}} + \Gamma_{\mathrm{corotation}}
.\end{equation}

A positive torque exerted by the disk on the planet means that the disk gives angular momentum to the planet and therefore that the planet migrates outward. In contrast, a negative total torque results in the disk gaining angular moment from the planet, and the planet migrating inward. Thus, at the locations where the total torque sign changes, we can define two physical radii:
\begin{itemize}
\item If the total torque is negative inside and positive outside that line, the potential planets are driven away from that location, therefore defining a depleted region in planetary cores, corresponding to the planet deserts of \citet{hasegawa12}.
\item If the total torque is positive inside and negative outside, the potential planets converge toward this location, which we thus call a planetary trap. Such a zero-torque radius was called by \citet{lyra10} the equilibrium radius. \citet{hasegawa111} studied the vertical effects of the planet mass, but only considered fixed surface mass density radial profiles, while \citet{hasegawa112} estimated planet trap locations due to Lindblad torques alone.
\end{itemize}

\subsubsection{Density and temperature features at the origins of the traps and deserts}

\begin{figure}[htbp!]
\center
\includegraphics[width=\hsize]{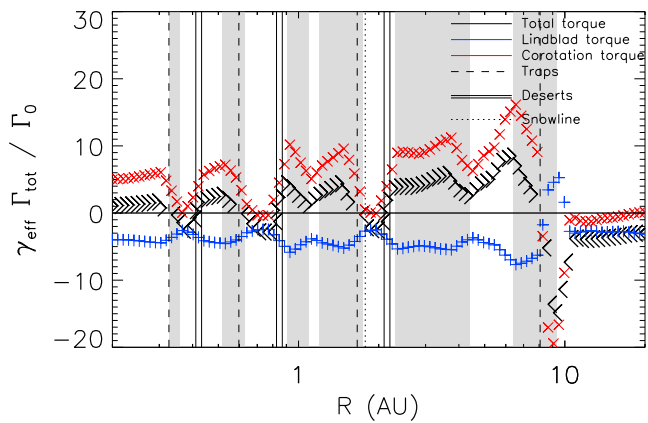}
\caption{Radial radial profile of the migration torques (black arrows showing the direction of migration: outward for positive torques and inward for negative torques) after 1 Myr of evolution. Lindblad torques (blue), corotation torques (red), and shadowed regions (gray) are also represented.}
\label{multitorque1M}
\end{figure}

Figure \ref{multitorque1M} presents the radial profiles of the Lindblad, corotation and total torques after 1 million years of evolution, when the planetary cores can already exist but the gas disk is not yet completely photo-evaporated. We chose to focus on the planetary formation region and therefore to limit the radial extent of our migration investigation up to 20 AU. The torques are normalized by $\Gamma_{0}$. A left-pointing arrow shows a negative torque and therefore an inward migration, while a right-pointing arrow shows a positive torque or outward migration.

\begin{figure}[htbp!]
\center
\includegraphics[width=\hsize]{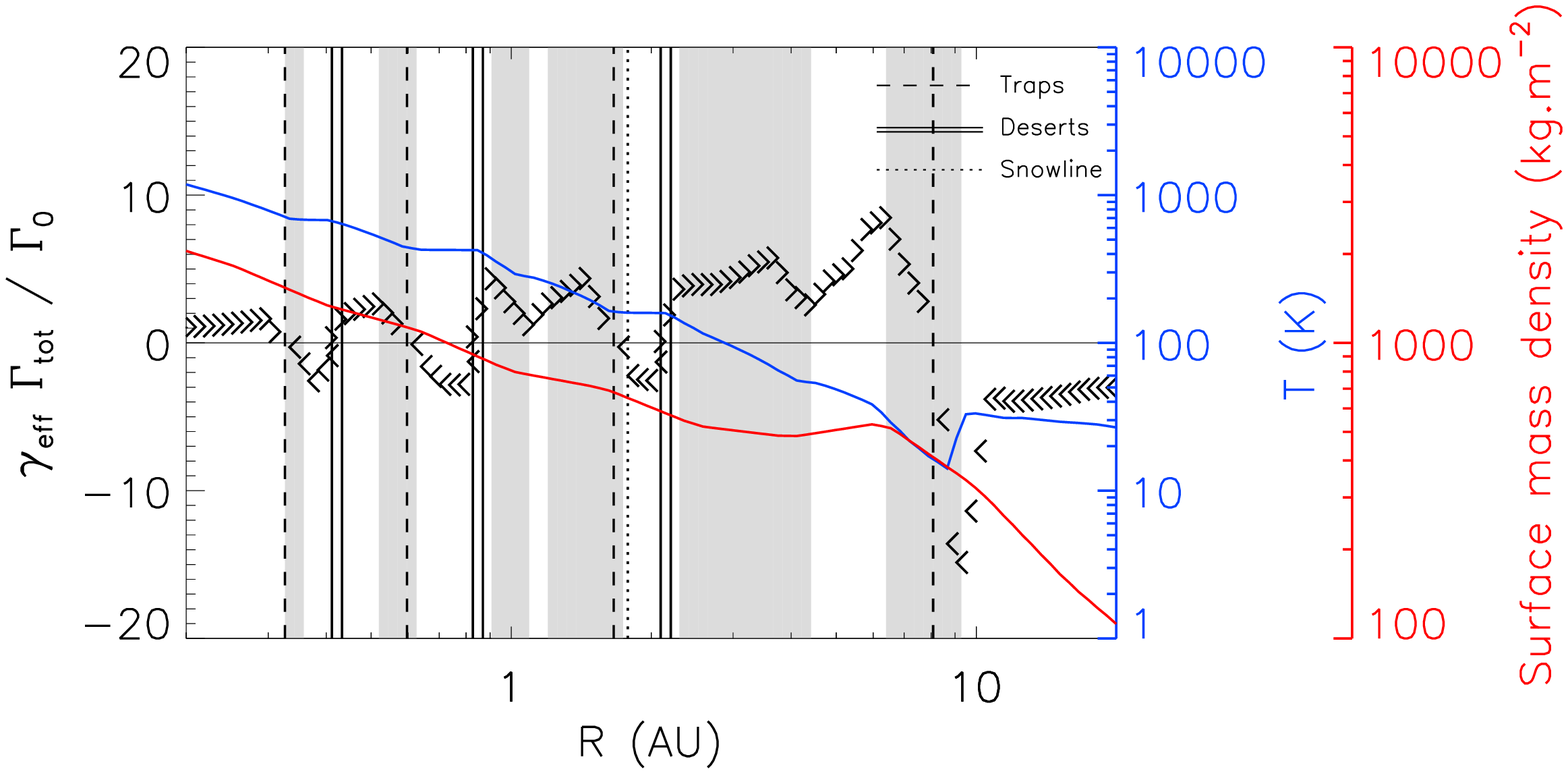}
\caption{Radial profile of the total torque (black arrows showing the direction of migration) after 1 Myr of evolution. Temperature radial profile (blue curve), surface mass density radial profile (red curve), and shadowed regions (gray regions) are also represented.}
\label{tottorque1M}
\end{figure}

The Lindblad torque is mostly negative except around 10 AU, where the temperature gradient is reversed. The corotation torque is mainly positive, but it can become negative in the temperature plateaux regions where the temperature gradient is shallower (Fig. \ref{tottorque1M}). The orders of magnitude of the two contributions are comparable, and the resulting total torque alternates between positive (outward) regions and negative (inward) regions. After 1 million years of evolution, outward migration could occur below 0.33 AU, between 0.41 and 0.60 AU, between 0.85 and 1.7 AU, and between 2.2 and 8.0 AU. As a consequence, planets could accumulate at the traps located around 0.33, 0.60, 1.7, and 8.0 AU and the regions around 0.41, 0.85, and 2.2 AU should be strongly depleted in planets. The snowline (dotted line at 1.8 AU in Fig. \ref{multitorque1M}) seems to be closely related to the nearby trap at 1.7 AU: the inner edge of the 160 K-plateau coincides with a trap, while there is a planetary desert at the outer edge of the plateau, as shown in Fig. \ref{tottorque1M},
which presents the total torque together with the surface mass density and temperature radial profiles. As a comparison, \citet{bitsch11} found a possible equilibrium radius around 12.5 AU for a 20 Earth-masses planet, which could be consistent with our 8 AU-trap.


As the total torque strongly depends on the surface mass density and temperature gradients (see Eqs. \ref{gammalin}-\ref{gammatot}), we can here relate the torque radial profile with the bumps in surface mass density and troughs in temperature. We note here that the density bump around 6 AU seems to result in a stronger positive torque gradient, in agreement with the estimate of \citet{masset06b} that sharp surface mass density gradients are required to slow-down type I migration. On the other hand, the temperature irregularity located between 9 and 10 AU seems to be at the origin of the torque inversion in this region. The density profile there does not seem to affect the total torque trend. This sharp temperature gradient appears to be correlated with an inner shadowed region, as expected by \citet{kretke12}, who described a model in which the absence of stellar irradiation could lead to an outward migration, and therefore estimated that sustaining an outward migration in an irradiated and active disk would be difficult without any shadowing effects. This correlation confirms the importance of properly taking into account the geometry and photosphere irradiation. \citet{kretke12} found traps inside 1 AU for the most massive embryos and estimated that a mass accretion rate $\dot{M}\, > \, 10^{-8}\, M_{\odot}.\mathrm{yr}^{-1}$ was required to trap planetary cores at radii compatible with known gas giant planet radii. This is compatible with our present study and the mass accretion rates obtained by \citetalias{baillie14} for evolution ages younger than 1 Myr.

\subsubsection{Earlier time evolution}

\begin{figure}[htbp!]
\begin{center} $
\begin{array}{cc}
\includegraphics[width=\hsize, clip=true]{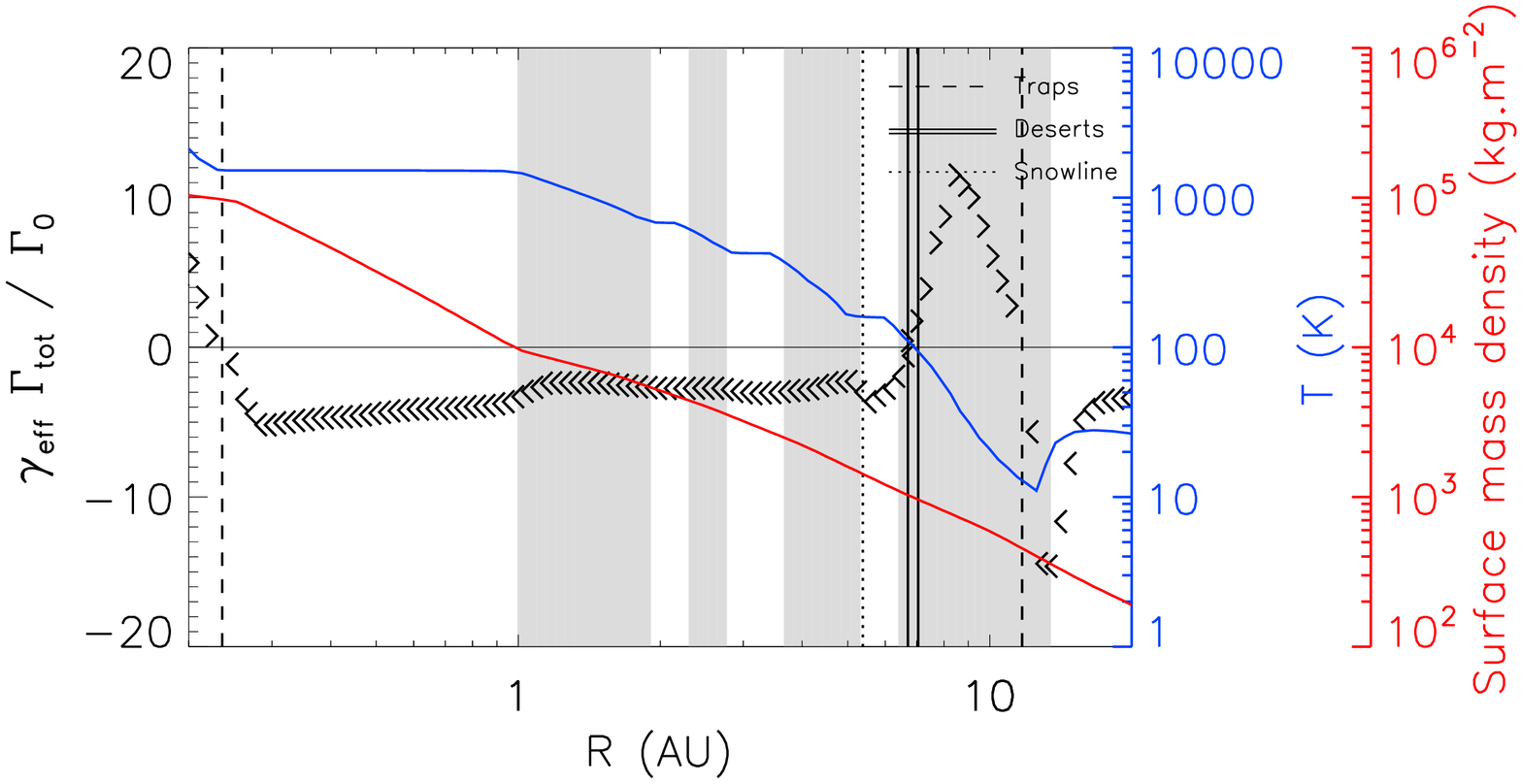}\\
\includegraphics[width=\hsize, clip=true]{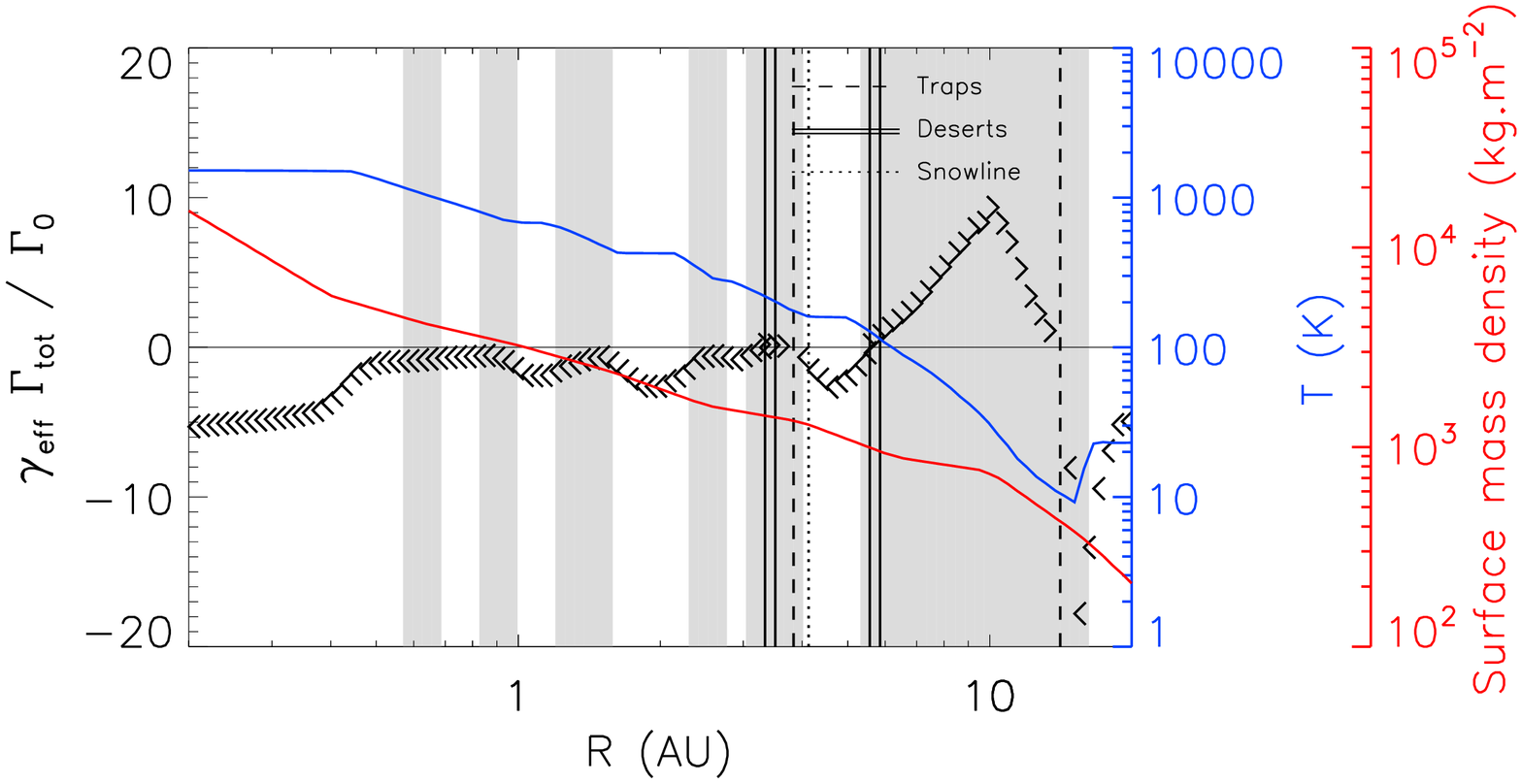}\\
\includegraphics[width=\hsize, clip=true]{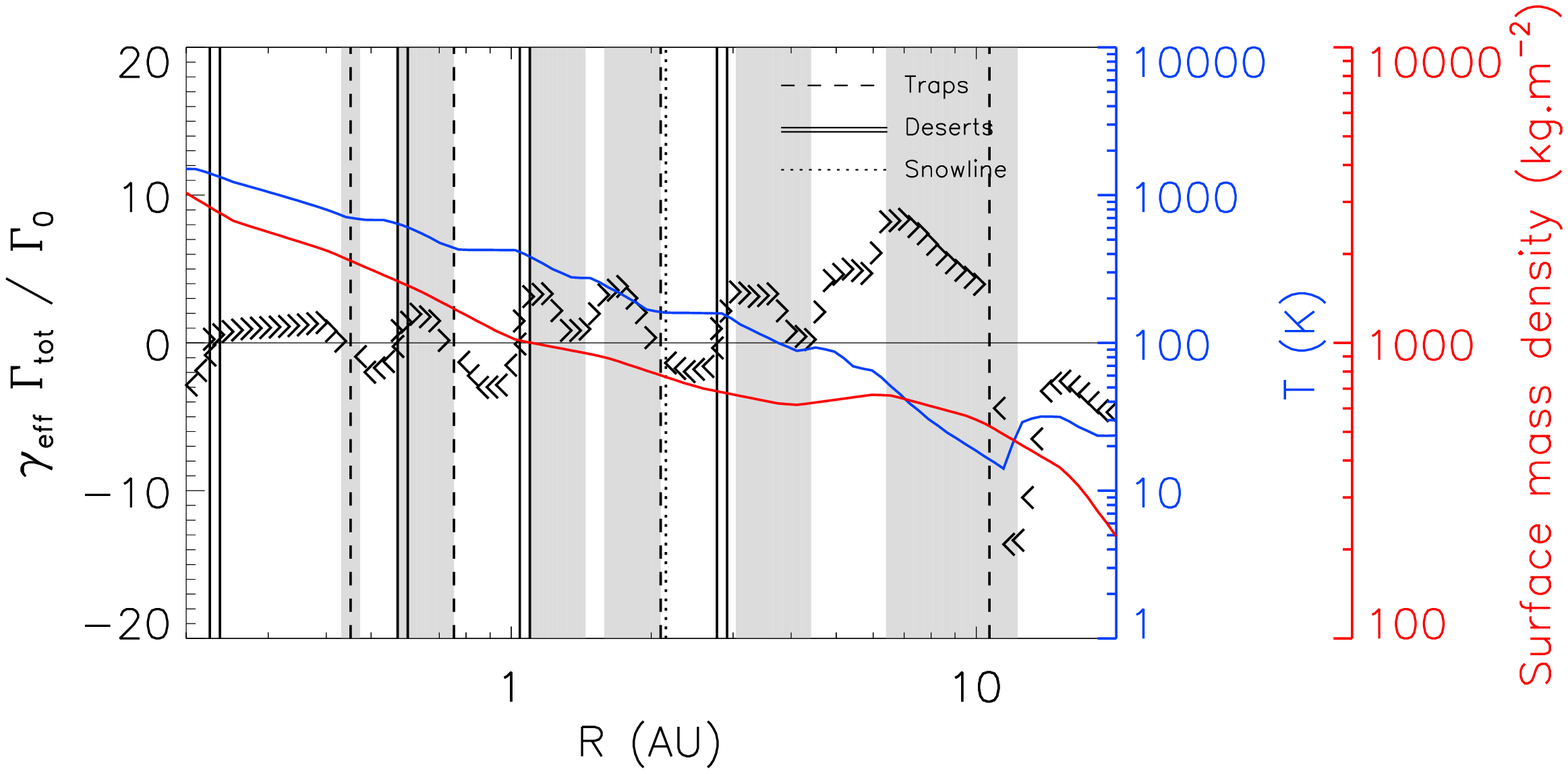}\\
\end{array} $
\end{center}
\caption{Radial profiles of the total torque (black arrows showing the direction of migration) after 1000 years (upper panel), 10,000 years (middle panel), and 100,000 years (lower panel) of evolution. Temperature radial profiles (blue curve), surface mass density radial profiles (red curve), and shadowed regions (gray regions) are also represented.}
\label{torquemultit}
\end{figure}

To follow the evolution of the trap and desert locations, Fig. \ref{torquemultit} presents the total torque profiles as functions of the radial distance for different ages. After 1000 years, we can only find two traps around 0.22 and 11 AU and one desert located at 7 AU. The trap is correlated with the shadow region just outside the temperature plateau at the water ice sublimation temperature (160 K).

After 10,000 years, we now have two traps (4 and 13 AU) and three deserts (3.5 and 6 AU), while the water ice sublimation plateau appears to be located around the 5 AU. The outer trap may be due to the temperature gradient inversion there.

After 100,000 years of evolution, we now have four traps (0.45, 0.75, 2, and 10 AU) and four deserts (0.23, 0.6, 1, and 3 AU). The traps appear to be located in the outer parts of the shadowed regions, while three of the deserts are at the inner edges of the shadowed zones. In addition, the water ice line coincides with the trap at 2 AU.

Traps and deserts seem to relate better to the variation in the temperature gradient than to the surface mass density gradient trends, as expected from the coefficients in Eq. \ref{gammalin}.

\subsubsection{Trap and desert migration}

\begin{figure}[htbp!]
\center
\includegraphics[width=\hsize]{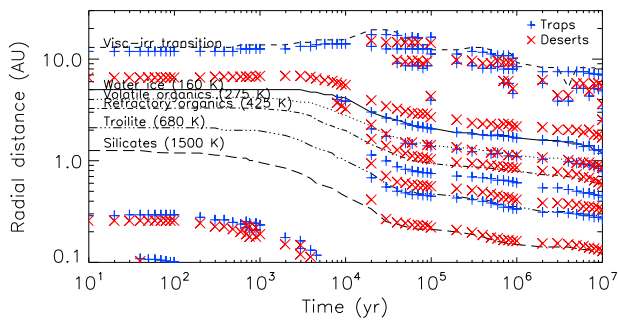}
\caption{Time evolution of the migration traps (blue "+") and deserts (red "x"). The snowline position (dotted line) and the heat transition radius (dashed line) are also represented.}
\label{traptime}
\end{figure}

To better follow the evolution of the traps and deserts, Fig. \ref{traptime} presents the variations in number and position of these traps and deserts.

Figure \ref{traptime} shows the evolution of the traps and deserts. We first note that in the early evolution of the disk, it seems to be possible to create traps and deserts below 0.3 AU. However, as these traps disappear in a few thousand years (much before a steady-state is reached), they probably only exist as long as the disk "remembers" the initial condition of the simulation. Second, it appears that there is a permanent trap correlated with the heat transition barrier (dashed line): this trap is systematically slightly (less than 1 AU) inward of that viscous-irradiation transition (consistent with the planetary trap position of a 20 Earth-masses planet estimated by \citet{bitsch11}), recalling
the estimate of \citet{hasegawa112} that the heat transition barrier was at the origin of trapping regions. Between 10,000 and 100,000 years, the heat transition barrier moves slightly outward, which may coincide with the outward drift of the relaxation radius that crosses the planetary formation region at this age (see Sect. \ref{sigma}). That trapping location is the outer boundary of an outward-migrating region that initially spreads from around 6.5 AU (about 1 AU outer to the water ice sublimation line) up to $\sim$ 13 AU. After 10,000 years of evolution, that outward-migrating region is divided into two (sometimes three) sequences of (outward-migrating + inward-migrating) regions. A depleted zone in planets can be found $\sim$ 1 AU outside of the water ice sublimation line at any time.

After 10,000 years, we also note traps and deserts following the main dust component sublimation lines:
\begin{itemize}
\item the water ice sublimation line coincides with a systematic trap,
\item the volatile organics sublimation line is closely accompanied by an inner trap and an outer desert,
\item the refractory organics and troilite sublimation lines are similarly bordered, if not as closely as the volatile organics line,
\item the silicate sublimation line is closely followed by an outer desert region.
\end{itemize}

From Fig. \ref{tottorque1M}, we can infer that the space between two sublimation regions is a zone of outward migration: therefore the chances of a planet to become trapped or significantly slowed down in the inner disk are not negligible. Once trapped, the forming planet is very likely to follow the trap migration. In addition, most of the planet traps (those that are correlated with the heat-transition barrier or the ice lines) seem quite sustainable: these traps and deserts slowly migrate inward with the sublimation lines as the disk ages and cools down. Only a few isolated transient traps and deserts appear between the water-ice line and the heat-transition barrier. However, it is very likely that a planet in such a disappearing trap would be trapped again in the next outer trap.

\citet{kley09} suggested that a possible scenario to address the inconsistency between the formation and migration timescales could be the retention of the icy cores at the snowline. We extend
this: it is possible to trap planets at any main dust component sublimation line. Dust sublimation is a physical mechanism sufficiently efficient to trap planets along the ice lines. In addition, as suggested by the correlation of the trap or desert locations with respect to the shadowed regions (Fig. \ref{tottorque1M}), the heating sources play an important role in trapping planets. This is reinforced by the presence of traps along the heat-transition barrier.

\subsubsection{Perspectives}

Although multiple planet-systems seem to be the most common configuration, we here only studied the interaction of a single planet with the disk. \citet{cossou13} showed that multiple planetary cores tend to be trapped in chains of mean motion resonances. Therefore, a proper treatment of multiple planet interaction with the disk requires considering the dynamics of these planets in addition to the disk dynamics. Coupling our hydrodynamical approach with an N-body code requires a more thorough study and will be the object of a future paper. In addition, as we have detailed, our current model does not take into account the feedback of the planet on the disk, which is currently not affected by the planet. This might also be necessary for future studies.

We leave the consideration of dead zones for
a future paper because they probably generate density and temperature irregularities that have been shown to be favorable for creating planetary traps and deserts. Although the torque expressions detailed above show a limited dependency on the planet mass ($\propto M_{\mathrm{P}}^{2})$), this is kept for a subsequent review, along with the dependency on the disk mass and the mass accretion rate.

\section{Conclusions}
We have studied the viscous evolution of a protoplanetary disk, using an $\alpha$ prescription for its viscosity. Starting with an initial minimum mass solar nebula, we self-consistently calculated the geometry, thermodynamics, and density distribution evolution of such a disk, taking into account the various sources of heating (viscous and irradiation), the shadowing effects, and the physical composition. The physical phases of the various dust components have a dramatic influence on the opacity of the disk and therefore reflect on the mid-plane temperature. The evolved density radial profiles present a few irregularities compared to the previous
results of \citetalias{baillie14}. These density bumps mainly result in temperature features (bumps, troughs, plateaux) and also widen the snowline and sublimation line regions. Our main result is that snowlines may better be called snow regions, AU-wide smooth variations, rather than sharp frontiers in the disk. Close to the sublimation temperature of the main dust elements, the disk mid-plane temperature becomes constant. This could result in a less sudden and strong change of phase than initially thought, involving partial sublimation. Given an element in Table \ref{tempchgt}, and if we focus on the disk mid-plane region for which the temperature corresponds to that element sublimation temperature within a few Kelvin, the gas-to-dust ratio of that element would vary from $\sim 0$ at the outer edge of the temperature plateau to $\sim 1$ at the inner edge (the rest of the material is unaffected at these temperatures). A planetesimal crossing such a region inward might then be able to melt layer by layer, with possible consequences on its mineral petrology in the case of a future recondensation. Another way to model the partial sublimation would be to consider a vertical gradient of gas-to-dust ratio for the element in question: the element would be sublimated from the mid-plane up to a given altitude that would vary with the radial distance in the plateau. The farther into the plateau, the higher that limit, and the greater the proportion of the material of the column that is sublimated. Such a refinement would affect the dust vertical transport as studied by \citet{ciesla10b}, for example. Possible combinations of vertical and radial gradients of gas-to-dust ratio could also reflect the physics of the temperature plateaux.

From these density and temperature radial profiles, we  estimated the resonant torques that potential planetary cores would exert on the disk. From the total torque sign (resulting from the Lindblad and corotation resonances), we derived the locations of planetary traps and deserts and followed their own migration as the disk ages: after 1 Myr, four planetary traps were identified around 0.33, 0.60, 1.7, and 8.0 AU, with three planetary deserts around 0.41, 0.85, and 2.2 AU.

As stated by \citet{masset06b}, density and temperature discontinuities appear to be the key characteristics leading to the generation of planetary traps: temperature troughs, density bumps, changes of phases, and heat transitions clearly affect the potential planet migration. Previous works indicated that density and temperature irregularities can slow down type I migration to allow planets to grow in a reasonable time. Our evolution code shows that such structures naturally arise during the disk evolution when the coupling between the disk dynamics, thermodynamics, and geometry is taken into account: planets can be trapped at the main dust component sublimation lines. Additional simulations are still required to estimate by how much planetary embryos can be slowed down by being trapped, and whether they can actually migrate along with the planetary traps. It appears necessary to synchronize the planet and trap migration rates, and this study requires a more thorough calculation of the planet feedback on the disk. Other sources of discontinuities, such as deadzones, may affect these migration torques and create more planetary traps or deserts. This will be treated in a separate paper.

\begin{acknowledgements}
We thank Esther Taillifet for enlightening discussions. We are also indebted to Bertram Bitsch and Dominic Macdonald for valuable suggestions that improved the quality of the manuscript significantly. We also thank the referee for detailed and constructive comments that improved the quality of the paper. This work was supported by IDEX Sorbonne Paris Cit\'e. We acknowledge the financial support from the UnivEarthS Labex program of Sorbonne Paris Cit\'e (ANR-10-LABX-0023 and ANR-11-IDEX-0005-02).  
\end{acknowledgements}


\bibliography{bibliography}
\end{document}